\documentclass[preprint,prb]{revtex4}

\usepackage{amsmath}
\usepackage{amssymb}
\usepackage{graphicx}
\usepackage{bm}

\begin{document}

\title{Quantum-limited charge detection with two quantum point contacts}%
\author{Kang-Ho Lee$^1$}
\author{Kicheon Kang$^{1,2}$}
\email{kicheon.kang@gmail.com}
\affiliation{$^1$Department of Physics, Chonnam National University,
 Gwangju 500-757, Korea}
\affiliation{$^2$Institute of Experimental and Applied Physics, University
 of Regensburg, 93040 Regensburg, Germany}
\date{\today}

\begin{abstract}
We present a scheme for tuning two quantum point contacts as a
quantum-limited charge detector.
Based on the scattering matrix approach, we analyze a general
condition of quantum-limited
detection with a single-channel quantum detector possessing
time-reversal symmetry. From this
analysis we find that quantum-limited detection can be easily
realized with two quantum point
contacts connected in series, which is not possible if only a
single quantum point contact
is used. We also discuss
the sensitivity enhancement due to multiple reflections of the two point contacts.
\end{abstract}


\maketitle
\newcommand \tr {{\rm Tr}}

Detection of single charges in mesoscopic systems has generated a lot of interest
in recent years~\cite{Field93,Devoret00,Sprinzak02}.
On one hand, it is important for the operation of nano-scale devices and
solid-state realization of
quantum information processing with charge qubits.
On the other hand, charge detection inevitably causes back-action dephasing,
and the controlled dephasing provides an ideal playground for studying the
complementarity principle in quantum
theory~\cite{Bohr,Buks98,Sprinzak00,Khym06,Chang08}. A detector is called
quantum-limited (100$\%$ efficient) if the dephasing in the measured
system is only due to the information acquisition by the detector. A single
quantum point contact (QPC) is often used as a
detector of single
electrons~\cite{Field93,Buks98,Sprinzak02,Avinun-Kalish04,Chang08}.
However, a generic single QPC is found to have low efficiency because of
the large phase-sensitive contribution of dephasing which cannot be
read out by the detector
circuit~\cite{Kang05,Kang07,Chang08}. It has been
proposed that a quantum-limited detection (QLD) is possible with edge-state
interferometers at high magnetic field~\cite{Averin05,Pathak09}.
Time-reversal symmetry (TRS)
is broken in these detectors due to the high magnetic field.

In this Communication, we present a simple scheme of QLD with two QPCs
connected in series (Fig.~1) for detection of the charge state of a
nearby qubit.
The detector does not require a high external magnetic field, and preserves
the TRS.
Based on the scattering matrix analysis, we derive a general condition
of a QLD for a single-channel conductor with TRS.
In particular, we find that a QLD can be easily achieved by tuning
only the ratio of the reflection probabilities of the two QPCs.
This is important for constructing a high-efficiency detector and is
in strong contrast to the case
of a single QPC in the following context.
For a single QPC, strict mirror-reflection symmetry is required for
a QLD~\cite{Korotkov01,Pilgram02,Khym06a}.
However, the mirror-reflection symmetry cannot be controlled generically,
and it is the main reason
why a single QPC has a low efficiency~\cite{Kang05,Kang07}.
We also point out that the detection sensitivity can be improved due to
multiple reflections at the two QPCs.

A measurement of charge state is accomplished through the scattering processes
in a detector. We assume that the detector has only a single transverse channel at
zero temperature.
The scattering process is described by  the S-matrix
\begin{equation}
\mathbf{S}=
\left(\begin{array}{ccc}
r & t' \\
t & r' \\
\end{array} \right),
\label{eq:smatrix}
\end{equation}
which connects the two incoming and the two outgoing waves.
In the weak-coupling limit, the change of the scattering coefficients
due to an extra charge is small, and it can be described in terms of
a continuous weak measurement.
For an applied bias $V>0$ (that is, the electrons are injected from the left
lead of the detector), the measurement-induced ``disturbance" of the qubit state
is characterized by the dephasing
rate~\cite{Gurvitz97,Aleiner97,Hackenbroich98,Stodolsky99}
\begin{subequations}
\begin{equation}
\Gamma_d={eV\over 4h} {(\Delta R)^2\over R(1-R)}
 +{eV\over h}R(1-R){(\Delta\xi)^2},
\label{eq:deph-rate}
\end{equation}
where $R= |r|^2$ and $\Delta R$ are the reflection probability and
its change due to different qubit states, respectively.
$\xi = \arg(t/r)$ and $\Delta\xi$ are the relative phase of $t$ and
$r$, and its change due to different qubit states, respectively.
The first term of Eq.~(\ref{eq:deph-rate}) is equivalent to the rate
of measurement by the detector~\cite{Korotkov01,Pilgram02,Khym06a}.
On the other hand, for a reversed bias, $V<0$, $\Gamma_d$
is given by
\begin{equation}
\Gamma_d={e|V|\over 4h} {(\Delta R')^2\over R'(1-R')}
 +{e|V|\over h}R'(1-R'){(\Delta\xi')^2},
\end{equation}
where $R'=|r'|^2$
and $\xi' = \arg(t'/r')$.

The following constraints exist because of the unitarity of S-matrix:
\begin{equation}
 R=R', \,\, \xi + \xi' = \pi, \label{eq:relative phase}
\end{equation}
\end{subequations}
which presents the invariance of the measurement and dephasing rates
under reversal of the bias. This invariance might seem to be obvious, but
it is shown only by applying the unitarity of the scattering matrix.
%

To reach a QLD, the ratio of the measurement to the dephasing rates should be
maximized. This is achieved by the condition $\Delta\xi=0$ (or equivalently
$\Delta\xi'=0$). 
We find that, 
for a detector with TRS ($t=t'$), 
the QLD condition is reduced to
\begin{equation}
 f \equiv \arg(r/r') = \mbox{\rm const.} ,
\label{eq:qld}
\end{equation}
or $\Delta f=0$.
It is clear from this relation that the mirror-reflection symmetry ($r=r'$) is a
sufficient (not a necessary) condition~\cite{Note-mrs} for a QLD,
which was not possible
to achieve in practice with single QPCs~\cite{Kang05,Kang07,Chang08}.

Next, we analyze the QLD condition of two QPCs based on Eq.~(\ref{eq:qld}).
Each QPC (QPC$_a$ and QPC$_b$) is characterized by the scattering
matrix $S_\alpha$
($\alpha=a,b$):
\begin{equation}
\mathbf{S_\alpha}=
\left( \begin{array}{ccc}
r_\alpha & t_\alpha' \\
t_\alpha & r_\alpha' \\
\end{array} \right).
\end{equation}
The components of the S-matrix of the two QPCs in series are given by
\begin{eqnarray}
r &=& r_a + {{{t_a}{t_a'}{r_b}}\over{1-{r_a'}{r_b}}}, \;\;
 t = {{{t_a}{t_b}}\over{1-{r_a'}{r_b}}}, \nonumber \\
r' &=& r_b'+{{{t_b}{t_b'}{r_a'}}\over{1-{r_a'}{r_b}}}, \;\;
 t' = {{{t'_a}{t'_b}}\over{1-{r_a'}{r_b}}}.
\end{eqnarray}
%
%
We assume that only the first QPC (QPC$_a$) is interacting with the qubit. In this
case, $S_b$ is independent of the qubit state. We find the QLD condition $\Delta
f=0$ with $f$ of Eq.~(\ref{eq:qld}) given as
\begin{equation}
f = f(R_a,\theta,\phi) = \theta +
 \arg({{\sqrt{R_a}}-{\sqrt{R_b}e^{i\phi}}
 \over{\sqrt{R_b}-{\sqrt{R_a}}e^{i\phi}}})-\arg(r_b),
\end{equation}
where $R_\alpha=|r_\alpha|^2$ ($\alpha=a,b$),
$\phi=\arg(r'_ar_b)$ is the phase shift accumulated in one round-trip between
the QPCs, and $\theta=\arg(r_a)$~\cite{singularity}.

In the weak-coupling limit, the QLD condition of our setup can be approximated
as,
\begin{subequations}
 \label{eq:df}
\begin{equation}
 \Delta f \simeq {\partial{f}\over\partial{R_a}}\Delta R_a
  + {\partial{f}\over\partial{\theta}}\Delta\theta
  + {\partial{f}\over\partial{\phi}}\Delta\phi = 0,
\end{equation}
where
\begin{eqnarray}
 {\partial{f}\over\partial{R_a}} &=& {R_b\sin{\phi}\over{1+g^2}}
   {\sqrt{{R_b}\over{R_a}}-2\cos\phi+\sqrt{{R_a}\over{R_b}}
   \over\{{2\sqrt{R_aR_b}-(R_a+R_b)\cos\phi}\}^2}, \\
 {\partial{f}\over\partial{\theta}} &=& 1, \\
 {\partial{f}\over\partial{\phi}} &=& {(R_a-R_b)\over{1+g^2}}
   {\{{2\sqrt{R_aR_b}\cos\phi-(R_a+R_b)}\}
   \over\{{2\sqrt{R_aR_b}-(R_a+R_b)\cos\phi}\}^2},
\label{eq:partial}
\end{eqnarray}
\end{subequations}
with $g=g(R,\phi)={(R_a-R_b)\sin\phi\over{2\sqrt{R_aR_b}-(R_a+R_b)cos\phi}}$.
In general, we cannot easily find what is going on with this complicated condition.
In the following, we discuss
two simple examples (Fig.~1(a) and (b)) where the QLD condition can be
derived.

In practice, it is useful
to consider the general case of Fig.~1(a), because interactions with the qubit
are expected to be stronger in this case compared to that of Fig.~1(b).
In order to understand how a QLD is possible in this configuration, we use
a potential shift model (Fig.~2)~\cite{Kang07} for QPC$_a$.
In this model, the interactions with the qubit are taken into account by
the two parameters, $\Delta V$ and $\Delta x$.
That is, interactions of QPC$_a$
with an extra charge in the qubit induces
a change of the peak height ($\Delta V$) as well as
a parallel position shift ($\Delta x$) in the potential.
This can be expressed as
\begin{equation}
 V_1(x) = V_0(x-\Delta x) + \Delta V ,
\label{eq:potential}
\end{equation}
where $V_0(x)$ ($V_1(x)$) stands for the 1D potential of the QPC$_a$ in the absence
(presence) of an extra charge at the qubit. The two parameters $\Delta V$ and
$\Delta x$ represent the symmetric and the asymmetric responses to the
extra charge, and provide the current-sensitive and the phase-sensitive
contributions to dephasing, respectively.

The general QLD condition derived in Eq.~(\ref{eq:df})
can be applied to the special case under consideration.
We are mainly interested in the limit where a single
QPC has low detector efficiency.
In a single QPC with low efficiency, $\Delta R_a$ induced by $\Delta V$
is negligible
in Eq.~(\ref{eq:df}). In fact, this is
the case for generic single QPCs~\cite{Kang05,Kang07,Chang08}.
In this limit, the QLD condition is reduced to
$ \Delta\theta
  + {\partial{f}\over\partial{\phi}}\Delta\phi = 0$, which can be
achieved by controlling the parameters $R_a, R_b$, and $\phi$.
With the potential shift model (neglecting $\Delta V$), 
the two reflection amplitudes $r_a$ and
$r_a'$ are modified by the parallel shift $\Delta x$ as~\cite{Kang07}
\begin{equation}
 r_a \rightarrow r_a e^{2ik\Delta x}, \;\; r_a' \rightarrow r_a' e^{-2ik\Delta x},
\end{equation}
where $k$ is the wave number of the injected electrons.
Therefore, we get a simple relation $\Delta\arg(r_a) =
-\Delta\arg(r_a')$, or
equivalently, $\Delta\theta=-\Delta\phi$. This leads to the QLD condition of
\begin{subequations}
\begin{equation}
 \frac{\partial f}{\partial\phi} = 1 .
\end{equation}
We find that this condition is achieved if
\begin{equation}
\cos{\phi}=\sqrt{{R_a}\over{R_b}}.
\label{eq:qld-psm}
\end{equation}
\end{subequations}
This means that a QLD can be easily controlled by tuning only the
reflection probabilities of the QPCs with $R_a<R_b$ for a given
value of phase $\phi$. This is in contrast to the special QLD condition
of a strict mirror-reflection symmetry ($r=r'$) in the detector potential.

We have shown that two QPCs in series have crucial advantages for high-efficiency
detection compared to a single QPC. In the following, we also
discuss the sensitivity enhancement due to the existence of the 2nd QPC.
The sensitivity is related to the dephasing rate, since the latter represents
the maximum possible rate of charge detection which can be obtained in
principle. Here we compare the dephasing rate of a quantum-limited detector of two QPCs to that of a single QPC.
For a comparison, we again use the potential shift model of
Eq.~(\ref{eq:potential})
and neglect the contribution from $\Delta R_a$.
For a quantum-limited detector with two QPCs, we find
\begin{equation}
 \Gamma_d = \frac{eV}{h} R_a (1-R_a) \frac{1-R_b}{(1+R_aR_b-2R_a)^2} (\Delta\phi)^2,
 \label{eq:deph-2qpc}
\end{equation}
with the condition of Eq.~(\ref{eq:qld-psm}).
On the other hand, the dephasing rate of a single QPC is
\begin{equation}
 \Gamma_d = \frac{eV}{h} R_a (1-R_a) (\Delta\phi)^2.
 \label{eq:deph-1qpc}
\end{equation}
Note that the dephasing rate of Eq.~(\ref{eq:deph-2qpc}) is equivalent to the
actual measurement rate. In contrast, the dephasing rate of a single QPC (Eq.~(\ref{eq:deph-1qpc})) can be transformed
to the actual measurement rate only when an additional interferometer 
	is introduced with proper tuning of its parameters~\cite{Averin05}.
In any case, we find that quantum-limited detector with two QPCs is more sensitive
than a single QPC detector, for the parameter range
$\frac{1-\sqrt{1-R_b}}{2-R_b}<R_a<1$.
Therefore, one can always tune the two QPCs to make them more sensitive than a single QPC.
The enhancement of the sensitivity becomes more prominent as $R_b\sim R_a$ with $1-R_b\ll1$. Together with the condition of the QLD (Eq.~\ref{eq:qld-psm}),
our study shows how an
efficient and sensitive charge detector (which are the two main properties
of a `good' detector) can be achieved by using two QPCs.

Next, we briefly discuss the simpler case of Fig.~1(b), where only
the phase shift $\phi$ is affected by the qubit state,
while $\Delta R_a=\Delta\theta = 0$.
In this case, the QLD condition is reduced to $\frac{\partial f}{\partial\phi}=0$.
We find that it can be achieved if
\begin{equation}
 R_a = R_b \,.
\end{equation}
Therefore a QLD can be realized simply by setting the two
reflection probabilities equal. As in the case of the potential shift model for
the setup of Fig.~1(a),
this is a much easier constraint for
realization than the general QLD condition of $r=r'$.
The dephasing rate in the QLD limit of this setup is given by
\begin{equation}
 \Gamma_d =
 {eV\over{2h}}{{{R_a{(1-R_a)^2}}(1+\cos\phi)}\over{(1+{R_a^2}-2R_a\cos\phi)^2}}
 (\Delta\phi)^2.
\label{eq:deph-b}
\end{equation}
It is straightforward to show that the dephasing rate can be tuned to be
larger than that of a single QPC.

Aside from the issue of realizing a ``good" charge detector
(efficiency and sensitivity),
a peculiar point exists at $\phi=\pi$
in the dephasing rate of Eq.~(\ref{eq:deph-b}).
The dephasing rate vanishes at $\phi=\pi$,
despite the charge sensitivity of the detector.
In other words, the phase coherence of the qubit is fully restored even
though there are Coulomb interactions between the `detector' (conductor) and
the `system' (qubit). This can
be regarded as a demonstration that the acquisition of the information, rather
than the interaction-induced momentum kick, plays the
central role in the quantum mechanical complementarity~\cite{Khym06,Chang08}.
The interaction itself does not destroy the quantum coherence of the qubit
unless (part of) the state information is transferred to the detector as
a result of the interaction.

In conclusion,
we have proposed that two quantum point contacts connected in series
can be easily tuned
for quantum-limited charge detection. This is of practical importance because
a single quantum point contact has generically low efficiency and
there is no systematic way to improve
the efficiency of single quantum point contacts.
We have also discussed
the sensitivity improvement due to multiple reflections of the two point contacts.
Our study demonstrates that two quantum point contacts connected in series
have crucial advantages for realizing a good charge detector.

This work was supported by the
National Research Foundation of Korea under Grant No.~2009-007259 and
No.~2009-0084606, and by the LG Yeonam Foundation.


\begin{figure}[b]
\includegraphics[width=2.5in]{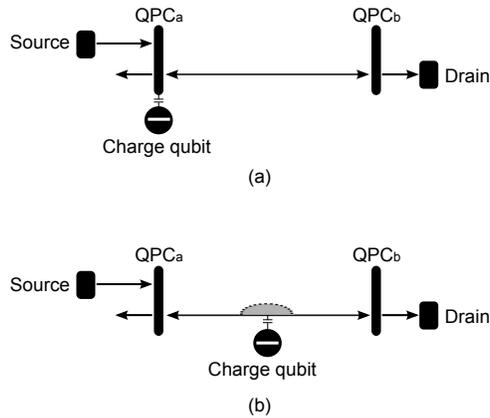}
\caption{\label{fig1}
Schematic for the detection of charge states in single charge qubits
with two QPCs connected in series. In setup (a), the first QPC interacts
with the qubit. In (b), the qubit is not directly interacting with the QPCs, but
induces a phase shift $\Delta{\phi}$ (shaded region) in the middle of the path.
}
\end{figure}

\begin{figure}[b]
\includegraphics[width=2.5in]{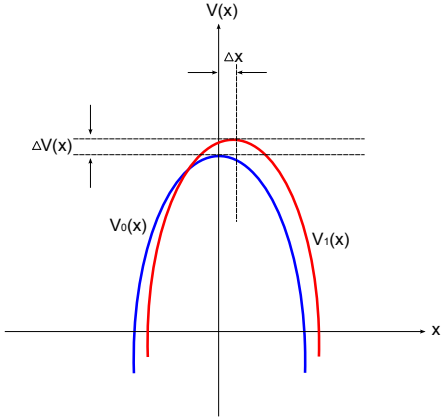}
\caption{\label{fig2}
A simple model for the potential response of QPC$_a$ due to the
interactions with the qubit. $V_i(x)$ stands for the QPC potential where the
qubit state is $i(\in{0, 1})$. The symmetric and the asymmetric responses
to the extra charge qubit are taken into account
via the parameters $\Delta{V}$ and $\Delta{x}$, respectively.
Note that no specific potential form is assumed.}
\end{figure}


\begin{thebibliography}{99}
\bibitem{Note-mrs} To be precise, the mirror-reflection symmetry is neither
 a sufficient condition for a QLD if resonant transport is involved.
 See Ref.~\onlinecite{Khym06a}.
\bibitem{singularity}
$\arg({{\sqrt{R_a}}-{\sqrt{R_b}e^{i\phi}}\over{\sqrt{R_b}-{\sqrt{R_a}}e^{i\phi}}})$
is indeterminate for $R_a=R_b$ and $\phi=0$. This case corresponds to a perfect
transmission ($R=0$) where abrupt phase change takes place. We exclude this
special point in our discussions.
\end{thebibliography}
\end{document}